\documentclass[acmlarge]{acmart}

\AtBeginDocument{%
  }

\acmJournal{IMWUT}
\acmVolume{0}
\acmNumber{0}
\acmArticle{0}
\acmYear{2026}
\acmMonth{9}

\usepackage{booktabs}
\usepackage{multirow}
\usepackage{subcaption}
\usepackage{xcolor}
\usepackage{soul}
\usepackage{enumitem}
\usepackage{tikz}
\usetikzlibrary{positioning, arrows.meta, fit, decorations.pathreplacing, calc}
\usepackage{pgfplots}
\pgfplotsset{compat=1.18}
\usepackage{xspace}

\newcommand{\sysshort}{\textsc{Pulse}\xspace}
\newcommand{\pulse}{\textsc{Pulse}\xspace}
\newcommand{\erdesire}{ER\_desire}  % Emotion regulation desire
\newcommand{\intavail}{INT\_availability}  % Intervention availability

\newcommand{\reviewrevision}[1]{{#1}}
\newcommand{\postreviewedit}[1]{#1}
\begin{document}

%% Title
\title{PULSE: Agentic Investigation with Passive Sensing for Proactive Affective Intervention in Cancer Survivorship}

\author{Zhiyuan Wang}
\email{vmf9pr@virginia.edu}
\affiliation{%
  \institution{Department of Systems and Information Engineering, University of Virginia}
  \city{Charlottesville}
  \state{Virginia}
  \country{United States}
}

\author{Subigya Nepal}
\email{sknepal@virginia.edu}
\affiliation{%
  \institution{Department of Computer Science, University of Virginia}
  \city{Charlottesville}
  \state{Virginia}
  \country{United States}
}

\author{Ariful Islam}
\email{xef4hb@virginia.edu}
\affiliation{%
  \institution{Department of Systems and Information Engineering, University of Virginia}
  \city{Charlottesville}
  \state{Virginia}
  \country{United States}
}

\author{Indrajeet Ghosh}
\email{vkw4ze@virginia.edu}
\affiliation{%
  \institution{Dept. of Systems and Information Engineering, University of Virginia}
  \city{Charlottesville}
  \state{Virginia}
  \country{United States}
}

\author{Xinyu Chen}
\email{dfs3mc@virginia.edu}
\affiliation{%
  \institution{Department of Systems and Information Engineering, University of Virginia}
  \city{Charlottesville}
  \state{Virginia}
  \country{United States}
}

\author{Katharine E. Daniel}
\email{ked4fd@virginia.edu}
\affiliation{%
  \institution{Center for Behavioral Health and Technology, University of Virginia}
  \city{Charlottesville}
  \state{Virginia}
  \country{United States}
}

\author{Laura E. Barnes}
\email{lb3dp@virginia.edu}
\affiliation{%
  \institution{Dept. of Systems and Information Engineering, University of Virginia}
  \city{Charlottesville}
  \state{Virginia}
  \country{United States}
}

\author{Philip Chow}
\email{pic2u@virginia.edu}
\affiliation{%
  \institution{Center for Behavioral Health and Technology, University of Virginia}
  \city{Charlottesville}
  \state{Virginia}
  \country{United States}
}

%% Abstract
\begin{abstract}
Cancer survivors face elevated rates of depression, anxiety, and emotional distress, yet self-report may be unavailable at some moments when support is relevant, a challenge we term the \emph{diary paradox}. We present \sysshort{}, a system for \emph{agentic sensing investigation}: LLM agents equipped with eight purpose-built tools query smartphone sensing data, compare current behavior with personal baselines, and retrieve outcome-labeled historical cases. Rather than receiving only a fixed feature summary, agents choose which modalities and time windows to inspect. We evaluate PULSE through a 2$\times$2 design crossing system architecture (structured single-pass vs.\ multi-turn agentic) with concurrent input modality (no current diary vs.\ sensing plus current diary) on 50 cancer survivors. The agentic multimodal condition achieves balanced accuracy of 0.743 for emotion-regulation desire; the agentic no-current-diary condition achieves 0.713 for self-reported intervention availability. This is a system-level comparison because the architecture conditions also differ in tool-mediated information access. The results provide a retrospective benchmark for interactive sensing investigation and motivate prospective evaluation at diary non-response moments.
\end{abstract}

%% CCS Concepts



%% Teaser figure
\begin{teaserfigure}
  \centering
  \includegraphics[width=\textwidth]{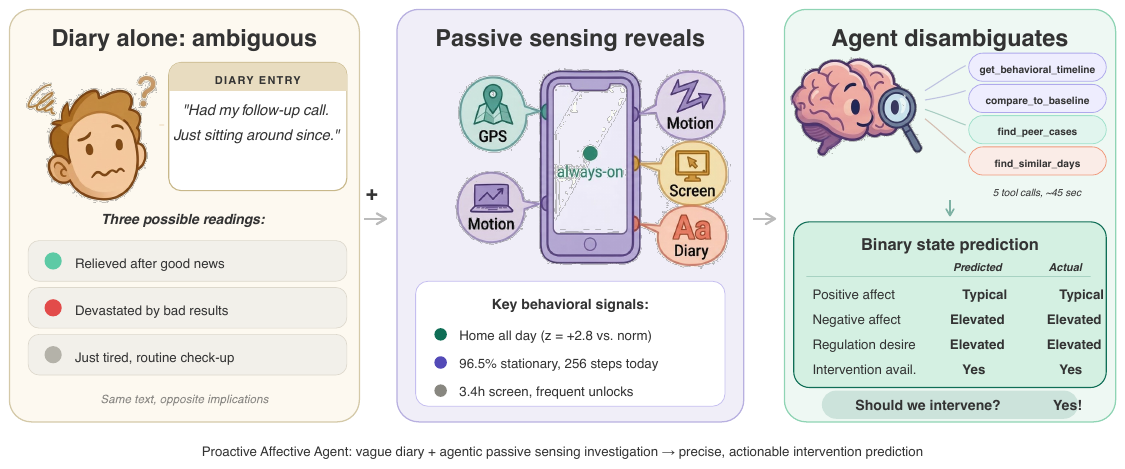}
  \caption{\postreviewedit{Conceptual illustration of the \sysshort{} interface. A brief diary can admit several interpretations (left); smartphone sensing supplies additional behavioral measurements (center); and an LLM queries purpose-built tools before returning study-target predictions (right). The figure does not represent a validated intervention decision rule.}}
  \Description{A three-panel figure showing the PULSE pipeline: (1) an ambiguous diary entry with three possible interpretations, (2) passive smartphone sensors including GPS, motion, screen, and diary, and (3) the PULSE agent using tools to produce binary state predictions for positive affect, negative affect, regulation desire, and intervention availability.}
  \label{fig:teaser}
\end{teaserfigure}

\maketitle

%% ================================================================
%%  1. INTRODUCTION
%% ================================================================
\section{Introduction}

\postreviewedit{More than 18 million cancer survivors live in the United States~\cite{acs2025survivorship}, and depression and anxiety are well-documented concerns in this population. ~\cite{mitchell2011depression}. Clinical guidelines recommend routine screening and management of these symptoms~\cite{andersen2023asco}, yet timely and scalable mental-health support remains difficult to provide. Just-in-time adaptive interventions (JITAIs) use mobile technology to adapt support across decision points to suit the needs of users ~\cite{nahum2018jitai}. Systems that target mental health may need to estimate both whether a person wants to regulate their current emotional state and whether they report being available to engage. In JITAI terminology, tailoring variables, intervention options, decision points, and decision rules jointly specify when, whether, and how support is offered~\cite{nahumshani2014jitai}.}

\postreviewedit{Ecological momentary assessment (EMA) provides repeated self-report through brief mobile surveys~\cite{shiffman2008ecological}. It is valuable as research ground truth and as a potential model input, but mobile-health missingness can be non-random; in some cohorts, affective states also predict subsequent prompt nonresponse~\cite{goldberg2021missing, murray2023ema}. We call the resulting design tension the \emph{diary paradox}: when a support system that depends on a completed current diary has no diary text to use at omitted prompts. CALLM~\cite{wang2026callm}, for example, uses diary text effectively when it is available. This motivates evaluating how well a system performs when the current diary is withheld.}

\postreviewedit{Passive smartphone sensing provides continuous, unobtrusive streams of behavioral data---movement patterns, location traces, screen usage, sleep timing, and communication proxies---that can accumulate without a contemporaneous diary response. A decade of research, from StudentLife~\cite{wang2014studentlife} and early digital phenotyping~\cite{saeb2015phenotyping, onnela2016harnessing} to GLOBEM~\cite{xu2022globem}, has established associations between these signals and mental-health outcomes. Generalization remains difficult: performance can approach chance in cross-dataset or population-level settings, although within-population personalization can perform substantially better~\cite{xu2022globem, adler2022machine, meegahapola2023generalization}.}

\postreviewedit{Why do sensing models often struggle outside the settings in which they were trained? Several factors are plausible, including sensor missingness, distribution shift, label noise, and limited representation of person-specific context. We study one of these factors: how behavioral evidence is assembled and interpreted for a particular person and decision point. Fixed pipelines apply the same feature transform to every example, even though an absolute value may mean very different things across people. A late-night screen session may be routine for one person but anomalous for another; a day of low mobility may matter differently after an active week than after a consistently sedentary one. An interactive system can request the evidence needed to make those comparisons and relate observations to personal history. We refer to this capability as \emph{contextual behavioral signal interpretation}~\cite{narayanan2013bsp}.}

Recent advances in large language models (LLMs) suggest a different path forward. LLMs can synthesize behavioral evidence in natural language, and prior work has shown that they can reason about both mental health text and wearable sensing summaries~\cite{kim2024health_llm, xu2024mental, xu2024penetrative_ai}. Most systems, however, still provide a pre-formatted summary whose contents are fixed before inference. The model cannot decide what to inspect, which anomaly to pursue, or how to compare today's behavior against the right personal baseline.

In this paper, we pose two questions: \emph{(1) how accurately do LLM-based systems predict the study's affect and availability targets with and without the current diary, and (2) how does a multi-turn, tool-mediated system compare with a structured single-pass system?}

\postreviewedit{We introduce \pulse{}, which gives an LLM eight sensing-query tools implemented through the Model Context Protocol~\cite{anthropic2024mcp}. Across multiple turns, the model selects modalities and lookback windows, requests personal-baseline comparisons, and returns a prediction. The prompt constrains the task and available tools but does not prescribe one fixed query sequence; the implementation follows a ReAct-style alternation between generated analysis and tool actions~\cite{yao2023react}. We compare this interface with a structured single-pass baseline.}

Figure~\ref{fig:teaser} gives a high-level view of how these sensing queries add behavioral context to an ambiguous diary entry.

We evaluate \pulse{} through a 2$\times$2 design crossing (1) \emph{system design}---structured (fixed summary, single LLM call) versus agentic (multi-turn tool use)---and (2) \emph{concurrent input}---no current diary versus sensing plus the current diary. This yields Struct-Sense, Auto-Sense, Struct-Multi, and Auto-Multi, evaluated retrospectively on 50 cancer survivors from the BUCS (Better Understanding Cancer Survivors) observational study and 4,112 completed EMA entries.

We define a candidate \emph{intervention opportunity} as the conjunction of two complementary constructs. \emph{Emotion regulation desire} measures whether the user wants to regulate their current emotional state, while \emph{intervention availability} records whether the user reports being available to engage with support. Modeling them separately preserves the distinction between a desire to change one's affect and the practical availability to do so.

Our contributions are:

\begin{enumerate}
    \item \postreviewedit{\textbf{A tool-mediated sensing investigation design.} We introduce LLM agents that query passive-sensing data through eight purpose-built MCP tools and compare the design with a structured summary pipeline. The architecture and tools can be extended to other passive-sensing domains.}

    \item \textbf{A 2$\times$2 evaluation of architecture and concurrent input modality.} We compare a structured single-pass pipeline with the agentic system under matched current-input conditions. \reviewrevision{This is a system-level comparison because the architectures differ in their interactive tool access and in the historical context available through that interface.}

    \item \textbf{Retrospective affect and availability prediction in a clinical population.} \pulse{} achieves 0.743 balanced accuracy on emotion-regulation desire when a current diary is available and 0.713 for self-reported intervention availability without a current diary.

    \item \postreviewedit{\textbf{A target-dependent current-diary association.} In this retrospective benchmark,  incorporating the current diary substantially improves emotion-regulation desire prediction (0.651$\rightarrow$0.743 BA) but has only a marginal effect on intervention availability prediction (0.713$\rightarrow$0.722 BA), motivating separate modeling of motivational state and stated availability. }
    
    %adding the current diary changes availability BA from 0.713 to 0.722 and emotion-regulation-desire BA from 0.651 to 0.743. This pattern motivates separate modeling of desire and stated availability.}
\end{enumerate}

The manuscript is organized as follows. Section~\ref{sec:paa-related} reviews related work at the intersection of passive sensing, LLMs for health, agentic AI, and JITAIs. Section~\ref{sec:paa-system} presents the \pulse{} system design, including the factorial framework, tool interfaces, and agent architectures. Section~\ref{sec:paa-evaluation} describes the evaluation methodology, and Section~\ref{sec:paa-results} presents results. Section~\ref{sec:paa-discussion} discusses implications and limitations, Section~\ref{sec:paa-future} outlines future work, and Section~\ref{sec:paa-conclusion} concludes.

% ===========================================================================
% 2. RELATED WORK
% ===========================================================================
\section{Related Work}
\label{sec:paa-related}

\pulse{} sits at the intersection of four research threads: passive mobile sensing for mental health, large language models for health and affective computing, agentic AI with tool use, and JITAIs.

% ---------------------------------------------------------------------------
\subsection{Passive Sensing for Mental Health}

\postreviewedit{The foundational insight that smartphone sensor data correlates with mental-health outcomes dates to StudentLife~\cite{wang2014studentlife}, which demonstrated associations between passively sensed behavioral patterns and depression, stress, and academic performance in college students. Saeb et al.~\cite{saeb2015phenotyping} further reported correlations between GPS features and depressive symptom severity, while Canzian and Musolesi~\cite{canzian2015trajectories} studied mobility traces for tracking depressive states. Onnela and Rauch~\cite{onnela2016harnessing} formalized ``digital phenotyping,'' and Torous et al.~\cite{torous2017new} connected it to the Research Domain Criteria framework. Subsequent work has examined relapse and symptom change in schizophrenia~\cite{wang2016crosscheck}, depression dynamics~\cite{wang2018tracking}, digital biomarkers of mood disorders~\cite{jacobson2019digital}, behavioral changes during the COVID-19 pandemic~\cite{huckins2020mental, nepal2022covid}, and depression onset from longitudinal passive data~\cite{chikersal2021detecting}. Instrumentation frameworks such as AWARE~\cite{ferreira2015aware}, long-running cohort studies~\cite{nepal2024college}, and recent reviews~\cite{dpmentalhealth2023review, dpstressanxiety2024} further characterize this literature.}

\postreviewedit{Alongside these advances, cross-context generalization remains a persistent challenge. GLOBEM~\cite{xu2022globem} evaluated 18 algorithms across four longitudinal sensing datasets and found that many cross-dataset depression-detection results were near chance. Meegahapola et al.~\cite{meegahapola2023generalization} likewise reported weak population-level mood inference across countries, while also showing that country-specific personalization can be much stronger. Adler et al.~\cite{adler2022machine} found that pooling longitudinal datasets sometimes helped but did not consistently solve transportability. These findings motivate careful separation of cross-population, within-population, and personalized evaluation rather than a single universal ``accuracy ceiling.''} \postreviewedit{As Adler et al.~\cite{adler2024beyond} argued in ``Beyond Detection,'' the field must connect detected correlates to actionable clinical support. One open design question is how to represent person- and episode-specific context. For example, 500 steps may have different meanings for users with very different personal baselines, and interpretation may depend on time of day, phone usage, location, and longitudinal history. \pulse{} explores one possible interface for assembling such context: an LLM can request selected behavioral summaries and personal-baseline comparisons through tools.}

% ---------------------------------------------------------------------------
\subsection{LLMs for Affective Computing and Health}

\postreviewedit{The application of LLMs to health prediction builds on few-shot prompting~\cite{brown2020language} and instruction following in aligned models~\cite{ouyang2022instructgpt}. Singhal et al.~\cite{singhal2023medpalm} evaluated clinical knowledge in LLMs, while Khasentino et al.~\cite{cosentino2025phllm} studied personal-health data including sleep and fitness metrics. Health-LLM~\cite{kim2024health_llm} evaluated 12 models on 10 wearable-sensing targets. Mental-LLM~\cite{xu2024mental} reported competitive depression- and stress-detection performance from online text. Zhang et al.~\cite{zhang2024leveraging} combined LLMs with structured smartphone-sensing summaries and reported gains over traditional ML in some configurations. SensorLLM~\cite{sensorllm2025} aligned LLMs with motion-sensor data for activity recognition, and a recent survey~\cite{llm_wearable_survey2024} reviews LLM applications for wearable health data. Zhang et al.~\cite{zhang2024sentiment} also document limitations relevant to LLM-based affect prediction.}

\postreviewedit{The most directly comparable published LLM baseline for the same population and targets is CALLM~\cite{wang2026callm}, which combines diary text, TF--IDF cross-user retrieval, and longitudinal memory. CALLM requires a current diary; \pulse{} additionally evaluates a condition in which the current diary is withheld.}

\postreviewedit{Several systems combine LLMs with sensing or personal-tracking data. MindScape~\cite{nepal2024mindscape} contextualizes journal prompts; GLOSS~\cite{choube2025gloss} uses generated code for open-ended sensing analysis; LENS~\cite{xu2025lens} aligns sensing data with language models; and Vital Insight~\cite{li2025vitalinsight} supports expert sensemaking through visualization and human-in-the-loop analysis. PHIA~\cite{phia2026} generates and executes code to answer retrospective wearable-health questions, while IoT-LLM~\cite{an2024iotllm} addresses general IoT data. Their tasks, populations, inputs, and evaluation protocols differ, so we use them for design context rather than performance comparison.}

% ---------------------------------------------------------------------------
\subsection{Agentic AI and Tool Use}

\postreviewedit{Agentic LLM systems combine multi-step generation with actions such as tool calls. Chain-of-thought prompting is an influential technique for eliciting intermediate reasoning~\cite{wei2022chain_of_thought}; ReAct~\cite{yao2023react} interleaves reasoning traces and actions. Toolformer~\cite{schick2023toolformer} studied self-supervised tool use, Gorilla~\cite{patil2023gorilla} studied API calling, and Voyager~\cite{wang2023voyager} used an embodied agent with a growing skill library. Separately, retrieval-augmented generation~\cite{lewis2020retrieval} supplies external examples or knowledge to generation, and a recent review summarizes its healthcare applications~\cite{gao2024ragsurvey}. \pulse{} combines these design patterns through behavioral-data tools and outcome-labeled retrieval.}

\postreviewedit{Wang et al.~\cite{wang2024agent_survey} survey LLM agents in terms of planning, memory, and tool use, and Huang et al.~\cite{huang2024planning} focus on planning. The Model Context Protocol~\cite{anthropic2024mcp} specifies an interface between models, tools, and data sources. ContextAgent~\cite{yang2025contextagent} studies proactive agents over wearable observations. ReflecTool~\cite{liao2025reflectool} studies reflection and tool verification on clinical tasks; its episode-level memory is conceptually related to \pulse{}'s reflection document, but the implementations and evaluations differ.}

LLMs can be unreliable numerical reasoners when asked to manipulate tabular data directly. \pulse{} therefore assigns counting, aggregation, thresholding, and temporal retrieval to deterministic tool code, while the model decides what to inspect and how to interpret the returned evidence.

\subsubsection{Agentic and LLM-Based Approaches in Health Sensing.}

\postreviewedit{Table~\ref{tab:paa-positioning} compares \pulse{} with several families of behavioral-health systems rather than presenting a linear progression. Conventional sensing systems such as GLOBEM~\cite{xu2022globem} and Mishra et al.~\cite{mishra2021receptivity} use engineered features and supervised models. LLM-based systems---including Health-LLM~\cite{kim2024health_llm}, Mental-LLM~\cite{xu2024mental}, Zhang et al.~\cite{zhang2024leveraging}, Feng et al.~\cite{feng2026comparative}, MindScape~\cite{nepal2024mindscape}, and CALLM~\cite{wang2026callm}---use preformatted sensor or diary representations for different tasks and populations. GLOSS~\cite{choube2025gloss} and PHIA~\cite{phia2026} allow an LLM to generate and execute analysis code for open-ended sensemaking or retrospective question answering. \pulse{} instead exposes domain-specific sensing queries and outcome-labeled retrieval for retrospective prediction of study-defined affect and availability targets. PH-LLM~\cite{cosentino2025phllm}, Time2Lang~\cite{pillai2025time2lang}, AMIE~\cite{tu2024amie}, and ReflecTool~\cite{liao2025reflectool} provide adjacent context but are not direct performance baselines.}

The systems also differ in what the model is allowed to do. GLOSS and PHIA give LLMs code-generation capabilities for exploratory analysis or retrospective questions, whereas ContextAgent~\cite{yang2025contextagent} reasons over wearable observations and ReflecTool~\cite{liao2025reflectool} studies reflection and tool verification for clinical tasks. \pulse{} uses a narrower action space: its tools expose predefined sensing summaries, temporal queries, personal comparisons, and retrieval operations. This restriction keeps aggregation and thresholding in deterministic code while allowing the model to choose which evidence to request. It trades the flexibility of arbitrary code generation for a reproducible interface tailored to repeated affect and availability prediction.

\begin{table}[htbp]
\centering
\caption{Positioning of \pulse{} relative to prior systems for health sensing and LLM-based health prediction. ``Domain Tools'' denotes whether the system provides purpose-built tools for autonomous investigation, as distinct from code generation or fixed prompting.}
\label{tab:paa-positioning}
\small
\resizebox{\textwidth}{!}{%
\begin{tabular}{lccccccc}
\toprule
\textbf{System} & \textbf{Year} & \textbf{Data Source} & \textbf{LLM} & \textbf{Agentic} & \textbf{Domain Tools} & \textbf{Clinical Pop.} & \textbf{Prediction} \\
\midrule
Mishra et al.~\cite{mishra2021receptivity} & 2021 & Sensing & No & No & No & No & Receptivity \\
GLOBEM~\cite{xu2022globem} & 2022 & Sensing & No & No & No & No & Affect/Depression \\
Health-LLM~\cite{kim2024health_llm} & 2024 & Wearable & Yes & No & No & Partial & 10 Health Tasks \\
Mental-LLM~\cite{xu2024mental} & 2024 & Text & Yes & No & No & Partial & MH from Text \\
Zhang et al.~\cite{zhang2024leveraging} & 2024 & Sensing & Yes & No & No & No & Affect \\
MindScape~\cite{nepal2024mindscape} & 2024 & Sensing+LLM & Yes & No & No & No & Journaling \\
CALLM~\cite{wang2026callm} & 2026 & Diary+RAG & Yes & No & No & Yes & Affect+Receptivity \\
GLOSS~\cite{choube2025gloss} & 2025 & Sensing & Yes & Partial & Code Gen. & No & Sensemaking \\
Feng et al.~\cite{feng2026comparative} & 2026 & Sensing & Yes & No & No & No & MH Forecasting \\
PHIA~\cite{phia2026} & 2026 & Wearable & Yes & Yes & Code Gen. & No & Retrospective Q\&A \\
\midrule
\textbf{\pulse{}} & \textbf{2026} & \textbf{Sensing+Diary} & \textbf{Yes} & \textbf{Yes} & \textbf{8 MCP Tools} & \textbf{Yes} & \textbf{Affect+Desire/Availability} \\
\bottomrule
\end{tabular}}%
\end{table}

Table~\ref{tab:paa-positioning} is therefore a map of design choices rather than a performance ranking. \pulse{} differs from the closest systems along three axes: it uses a multi-turn query loop rather than a single preformatted prompt; it exposes domain-specific sensing operations rather than unrestricted code execution; and it evaluates repeated study-target predictions in a cancer-survivor cohort rather than retrospective question answering or general-population sensemaking. These differences motivate the comparison but do not make scores across studies directly comparable.

% ---------------------------------------------------------------------------
\subsection{Just-in-Time Adaptive Interventions}

\postreviewedit{The JITAI framework~\cite{nahum2018jitai, nahumshani2014jitai} identifies six components: decision points, intervention options, tailoring variables, decision rules, proximal outcomes, and distal outcomes. Tailoring variables are measurements used to adapt delivery at decision points. A recent meta-analysis reported small average mental-health benefits alongside substantial heterogeneity and risk-of-bias concerns~\cite{jitai_effectiveness2025}.}

\postreviewedit{Kunzler et al.~\cite{kunzler2019receptivity} examined state-level receptivity to mHealth interventions using response and engagement measures. Mishra et al.~\cite{mishra2021receptivity} studied in-the-wild receptivity detection with smartphone and wearable context, including time, activity, and phone use. These operationalizations are related to, but not identical with, the self-reported desire and availability labels used here. Klasnja et al.~\cite{klasnja2015mrt} developed the microrandomized trial (MRT) methodology for evaluating JITAI components, later applied in HeartSteps~\cite{klasnja2019heartsteps}. IntelliCare~\cite{mohr2017intellicare} and MRT use notification timing~\cite{mello_emt2023} to provide some intervention context.}

\postreviewedit{\pulse{}'s emotion-regulation desire (\erdesire{}) and self-reported intervention availability (\intavail{}) targets can be treated as candidate tailoring variables. Gross's emotion-regulation framework~\cite{gross2015emotion} motivates the former broadly. The two study labels are related to, but distinct from, established receptivity constructs; we therefore analyze them separately and treat their joint use as a study-specific operationalization.}

Estimating a tailoring variable is only one part of JITAI design. A complete JITAI must also specify decision points, intervention options, decision rules, and proximal outcomes, and then test how those components affect behavior or well-being~\cite{nahumshani2014jitai, klasnja2015mrt}. \pulse{} focuses on the upstream state-estimation problem: how to assemble behavioral evidence when the available inputs vary across decision points. Prediction performance alone does not establish that acting on the estimate improves an intervention outcome.

% ===========================================================================
% 3. SYSTEM DESIGN
% ===========================================================================
\section{System Design}
\label{sec:paa-system}

\begin{figure}[t]
\centering
\includegraphics[width=\textwidth]{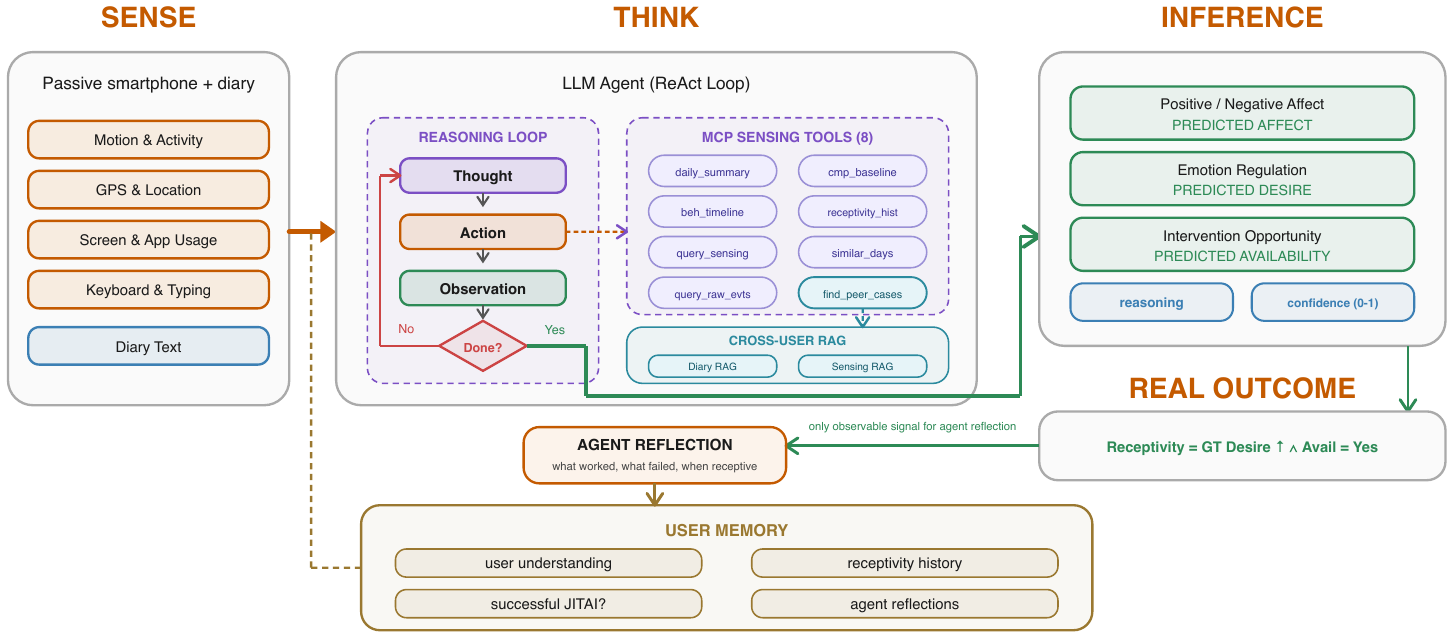}
\caption{System architecture of \pulse{}. Passive smartphone sensing streams and optional diary text (\emph{Sense}) feed an LLM agent that conducts a multi-turn ReAct investigation using eight domain-specific MCP tools and cross-user retrieval-augmented generation (\emph{Think}). The agent produces predicted affect, desire, and availability along with reasoning and confidence scores (\emph{Inference}). The baselines and 2$\times$2 design are described in Sections~\ref{sec:paa-factorial} and~\ref{sec:paa-baselines}.}
\Description{Architecture diagram with three stages: Sense assembles sensing streams and optional diary text; Think conducts a multi-turn investigation using tools and retrieval; and Inference outputs predicted affect, desire, availability, reasoning, and confidence.}
\label{fig:paa-architecture}
\end{figure}

Figure~\ref{fig:paa-architecture} presents the \pulse{} architecture: \emph{Sense} assembles smartphone sensing and, in multimodal conditions, the current diary; \emph{Think} uses a ReAct-style loop with eight MCP tools and outcome-labeled retrieval; and \emph{Inference} returns affect, desire, stated availability, a reasoning trace, and self-reported confidence.

% ---------------------------------------------------------------------------
\subsection{Problem Formulation}

\postreviewedit{Given the data assembled for a completed EMA entry $t$---which may include sensing history, the current diary, a user profile, and memory---\pulse{} predicts (1) three continuous targets: Positive Affect, Negative Affect~\cite{watson1988development}, and emotion-regulation desire; and (2) sixteen binary indicators, including individualized positive and negative affect states, discrete emotions, interaction quality, physical pain, future outlook, emotion-regulation desire state, and intervention availability.}

\postreviewedit{We focus on four study-defined, clinically relevant targets that could be evaluated as candidate JITAI tailoring variables~\cite{nahum2018jitai, mattingly2025predictors}. The two intervention-related targets are \textbf{ER\_desire}, which asks whether the user currently wants to regulate their emotional state, and \textbf{INT\_availability}, which asks whether the user reports being available to engage with an intervention. The affect targets, \textbf{PA\_State} and \textbf{NA\_State}, indicate whether positive or negative affect is high relative to the user's personal baseline.}

\postreviewedit{For this study, we operationalize the conjunction of ER\_desire and INT\_availability as a candidate intervention opportunity. This definition is a modeling choice, not a validated decision rule or a measure of actual intervention engagement.}

% ---------------------------------------------------------------------------
\subsection{Factorial Design}
\label{sec:paa-factorial}

\postreviewedit{To compare the agentic system across concurrent input modalities, we employ a 2$\times$2 design crossing two dimensions (Table~\ref{tab:paa-factorial}):}

\begin{table}[t]
\centering
\caption{The 2$\times$2 factorial design. Each cell represents a primary condition evaluated on the same 50 users. CALLM is evaluated as a diary-only baseline outside the factorial design.}
\label{tab:paa-factorial}
\small
\begin{tabular}{lcc}
\toprule
 & \textbf{No Current Diary} & \textbf{Sensing + Current Diary} \\
\midrule
\textbf{Structured} & Struct-Sense & Struct-Multi \\
\textbf{Agentic} & Auto-Sense & Auto-Multi \\
\bottomrule
\end{tabular}
\end{table}

\noindent\textbf{Reasoning Architecture.} \emph{Structured agents} (Struct-Sense, Struct-Multi) receive all available data as a pre-formatted summary and produce predictions in a single LLM call. Their fixed pipeline considers sleep, mobility, social signals, and peer comparisons before producing the final prediction. \emph{Agentic agents} (Auto-Sense, Auto-Multi) instead receive a user profile, optional diary text, and access to eight MCP tools. They choose what to investigate through a multi-turn tool-use loop, typically making 6--12 tool calls per prediction.

\noindent\textbf{Data Modality.} \reviewrevision{\emph{No-current-diary} conditions (labeled Struct-Sense and Auto-Sense for continuity)} omit the diary for the current prediction while \reviewrevision{still permitting temporally prior self-report outcomes through the historical channels described below.} \emph{Multimodal} conditions (Struct-Multi, Auto-Multi) provide both passive sensing data and the user's diary entry for that time point.

\noindent\textbf{Additional Condition.} We include one additional condition: CALLM~\cite{wang2026callm}, which uses diary text with TF-IDF cross-user retrieval and no sensing data, serving as a diary-dependent LLM baseline.

This design supports two primary comparisons. Within each concurrent-input condition, Auto-Multi vs.\ Struct-Multi and Auto-Sense vs.\ Struct-Sense compare multi-turn tool-mediated investigation with a structured single-pass pipeline. \reviewrevision{Because the interfaces expose different information, these contrasts compare the implemented systems; they do not isolate reasoning style alone.} Within each architecture, multimodal vs.\ no-current-diary conditions estimate the association with supplying the current diary. CALLM provides an additional diary-only reference outside the factorial design.

% ---------------------------------------------------------------------------
\subsection{The BUCS Dataset}

\pulse{} is evaluated on the BUCS dataset~\cite{mattingly2025predictors}, a longitudinal study of 407 adult cancer survivors who completed 5 weeks of data collection. The dataset was obtained in collaboration with the original study team and used with institutional review board approval. Participants completed three daily ecological momentary assessments (EMA)~\cite{shiffman2008ecological}, randomly delivered in the morning, afternoon, and evening. The surveys captured positive and negative affect ratings, discrete emotions, emotion regulation desire, intervention availability, and brief diary entries about emotional drivers.

\textbf{Passive smartphone sensing} continuously recorded accelerometer-based motion (stationary, walking, running, automotive activity), GPS location (distance traveled, time at home, location variance), screen usage (session counts, durations, lock/unlock patterns), app usage (category-level durations; Android only), and accelerometer-derived sleep periods. Keyboard activity (typing sessions and character counts) is sparsely available because participants could choose whether to share this modality. \textbf{User profiles} contain baseline demographic information, cancer type and treatment history, and validated psychological trait measures.

Platform differences between iOS and Android affect data availability: app-level usage is available only on Android (approximately 36\% of the test sample). The sensing tools are designed to gracefully handle missing modalities, returning explicit ``data not available'' responses that the agent incorporates into its reasoning. The data collection protocol was approved by the institutional review board (anonymized for review), and all participants provided informed consent.

% ---------------------------------------------------------------------------
\subsection{Structured Agents}
\label{sec:paa-structured}

The structured agents (Struct-Sense, Struct-Multi) implement a fixed pipeline in which all available data is pre-assembled and the LLM produces a prediction in a single call. This design mirrors the predominant approach in prior LLM-for-sensing work~\cite{zhang2024leveraging, kim2024health_llm}.

\textbf{Input Assembly.} For each prediction, the system pre-computes a sensing summary covering all available modalities for the current day \reviewrevision{up to the EMA timestamp} (sleep metrics, mobility statistics, screen usage patterns, activity breakdowns). This summary, along with the user's trait profile, a longitudinal memory document, and outcome-labeled peer examples, is assembled into a structured prompt.

\textbf{Cross-User Outcome Retrieval.} Both structured and agentic conditions receive outcome-labeled peer cases. \reviewrevision{Structured agents use TF--IDF text retrieval in multimodal mode or cosine similarity over daily sensing fingerprints in no-current-diary mode. The top 10 positive-similarity entries and their ground-truth EMA outcomes are included in the prompt.}

\textbf{Prompting Strategy.} Within a single LLM call, the prompt asks the model to consider the available behavioral modalities (sleep, mobility and activity, and social signals), reference the peer outcome examples, and produce a final prediction. The model's natural-language reasoning is open-ended, but its input context and output schema are fixed across predictions. \reviewrevision{The modality order is a fixed implementation choice and was not varied in this study.}

% ---------------------------------------------------------------------------
\subsection{Agentic Investigation Agents}
\label{sec:paa-agentic}

The agentic conditions (Auto-Sense, Auto-Multi) use a multi-turn loop that alternates generated analysis with tool actions, drawing on chain-of-thought prompting~\cite{wei2022chain_of_thought} and ReAct~\cite{yao2023react}. \reviewrevision{The model receives a task prompt, a user profile, a pre-generated memory document, access to eight MCP tools, and the current diary only in the multimodal condition. Tool selection is generated under the prompt's standing instructions, including mandatory peer retrieval. Appendix~\ref{app:prompts} documents the prompt structure.}

\subsubsection{MCP Tools.}
\label{sec:paa-tools}

\postreviewedit{A core design choice is to keep aggregation, retrieval, and temporal filtering in deterministic tool code while using the LLM for contextual interpretation. The tools support broad orientation (\texttt{get\_daily\_summary}, \texttt{get\_behavioral\_timeline}), targeted follow-up (\texttt{query\_sensing}, \texttt{query\_raw\_events}), within-person contextualization (\texttt{compare\_to\_baseline}, \texttt{find\_similar\_days}), and cross-user retrieval (\texttt{find\_peer\_cases}).}

The eight tools are implemented via the Model Context Protocol (MCP)~\cite{anthropic2024mcp}. Table~\ref{tab:paa-tools} summarizes their signatures and returned information. \reviewrevision{The tools use the following query windows and defaults: 0--7 prior days for \path{get_daily_summary}, 1--48 hours before the EMA for \path{query_sensing}, and 14 days for \path{get_receptivity_history}, matching the two-week recall period standard in affect measurement~\cite{kroenke2001phq}. \path{compare_to_baseline} returns a time-of-day-specific mean, dispersion, z-score, and qualitative bands at $|z|=.5,1,2$; \path{find_similar_days} searches pre-EMA same-user history and returns the top three days; and \path{find_peer_cases} defaults to ten positive-similarity cases from other participants. These bounds are design defaults matched to each tool's temporal granularity, in line with window lengths common in longitudinal mobile-sensing and EMA studies~\cite{wang2014studentlife,saeb2015phenotyping,mishra2021receptivity}; they were fixed a priori rather than tuned, and the agent chooses its actual lookback within them at query time. Retrieval depths follow the same pool-scaled logic: \path{find_peer_cases} returns 10--20 outcome-labeled peer cases as few-shot in-context examples~\cite{brown2020language,lewis2020retrieval}, a range validated in prior work on this population~\cite{wang2026callm} that balances inference gains against token cost, while \path{find_similar_days} returns three days from the far smaller
within-user history.}

\reviewrevision{Deterministic data-access code enforces the temporal boundary: sensing records, historical outcomes, memory, retrieval results, and structured summaries are restricted to information available before the current EMA timestamp. The current diary is supplied only in the multimodal conditions.}

\begin{table}[h]
\centering
% \color{blue}
\small
\begin{tabular}{@{}p{0.965\columnwidth}@{}}
\toprule
\textbf{\texttt{get\_daily\_summary}}\,(\texttt{date}, \texttt{lookback\_days}${=}0..7$)\\
\quad Natural-language summary of a day's behavioral patterns, with optional multi-day lookback.\\[3pt]
\textbf{\texttt{get\_behavioral\_timeline}}\,(\texttt{date}, \texttt{segment\_hours}${=}1..6$)\\
\quad Chronological reconstruction of the day before the EMA.\\[3pt]
\textbf{\texttt{query\_sensing}}\,(\texttt{modality}, \texttt{hours\_before\_ema}${=}1..48$, \texttt{hours\_duration}${=}1..24$, \texttt{granularity})\\
\quad Aggregates for one modality (GPS, motion, screen, keyboard, music, light) in a chosen window.\\[3pt]
\textbf{\texttt{query\_raw\_events}}\,(\texttt{modality}, \texttt{hours\_before\_ema}, \texttt{hours\_duration}, \texttt{max\_events}${=}30$)\\
\quad Raw event stream: app sessions, screen locks, activity transitions, typing sessions.\\[3pt]
\textbf{\texttt{compare\_to\_baseline}}\,(\texttt{modality}, \texttt{feature}, \texttt{hours\_before\_ema})\\
\quad Current value, time-of-day-specific historical mean/SD, z-score, and qualitative bands at $|z|=.5,1,2$.\\[3pt]
\textbf{\texttt{get\_receptivity\_history}}\,(\texttt{lookback\_days}${=}14$)\\
\quad Prior numeric regulation-desire scores and availability labels over the selected lookback.\\[3pt]
\textbf{\texttt{find\_similar\_days}}\,(\texttt{top\_k}${=}3$)\\
\quad Same-user past days by cosine similarity, with prior desire, availability, and diary snippets.\\[3pt]
\textbf{\texttt{find\_peer\_cases}}\,(\texttt{search\_mode}${=}$\texttt{text}$|$\texttt{sensing}, \texttt{top\_k}${=}10$ default; the prompt instructs 10--20)\\
\quad Outcome-labeled cases from \emph{other} participants (TF--IDF diary or sensing-fingerprint similarity); non-positive matches are omitted.\\
\bottomrule
\end{tabular}
\caption{\reviewrevision{The eight MCP tools used by the agentic conditions: signatures, parameter ranges, and returned information.}}
\label{tab:paa-tools}
\end{table}

\subsubsection{Investigation Loop.}

The \pulse{} agent operates through a multi-turn conversation with tool-use capabilities. A typical investigation begins with the user profile and any available diary text, followed by a broad view from \texttt{get\_daily\_summary} and \texttt{get\_behavioral\_timeline}. The agent can then issue targeted sensing queries, compare observations with the user's baseline, or retrieve similar days and peer cases before producing the final predictions, reasoning trace, and confidence estimate.

The prompt does not require this order. The agent may skip uninformative steps, revisit an earlier query after new evidence, or select a different investigation path. The maximum number of tool-use turns is capped at 16 to bound inference time, though most predictions complete in 6--12 turns (30--90 seconds).

\subsubsection{Per-User Session Memory.}
\label{sec:paa-memory}

As the agent processes sequential EMA entries for a user, it appends a brief reflection to a session-memory document. \reviewrevision{The reflection receives the investigation summary, the prediction, and a binary composite of elevated regulation desire and self-reported availability from the completed EMA. Prior desire and availability also enter through history tools and outcome-labeled peer cases, so the ablation in Section~\ref{sec:paa-memory-contribution} removes only the reflection document.}

% ---------------------------------------------------------------------------
\subsection{Outcome-Labeled Cross-User RAG}
\label{sec:paa-rag}

\pulse{} uses cross-user retrieval-augmented generation~\cite{lewis2020retrieval} to return cases with similar behavioral patterns or diary entries and their EMA outcomes. These cases provide outcome-labeled reference points for the model.

Retrieval uses TF--IDF similarity over diary text in multimodal conditions and cosine similarity over normalized daily sensing vectors in no-current-diary conditions. In the agentic conditions, \texttt{find\_peer\_cases} exposes both modes and the agent selects between them based on the available input.

The peer database is constructed from the 349-user training pool and excludes all 50 evaluation users.

% ===========================================================================
% 4. EVALUATION METHODOLOGY
% ===========================================================================
\section{Evaluation Methodology}
\label{sec:paa-evaluation}

% ---------------------------------------------------------------------------
\subsection{Evaluation Setup}

We evaluate \pulse{} on a test sample of 50 users randomly selected from the 399 BUCS participants with sufficient data (mean 82.2 entries per user, range 74--96). The remaining 349 participants form the training pool for trained baselines and cross-user retrieval. We chose 50 users as a balance between inference cost and coverage. Each agentic user consumes roughly seven million input tokens across ${\sim}$80 multi-turn predictions, making full-cohort evaluation of all 399 users across five systems prohibitively expensive. The main comparison includes the four 2$\times$2 conditions plus CALLM, each on 4,112 EMA entries for the 50 users. \reviewrevision{The 50 participants were aged 30--77 (M~=~53.5, SD~=~10.9), predominantly female (90\%) and White (94\%), with breast cancer the most common diagnosis (56\%) across stages from in-situ to stage~IV (Table~\ref{tab:app-demographics}, Appendix~\ref{app:demographics}).}

\subsubsection{Representativeness Analysis.}
\label{sec:paa-representativeness}

We compared the 50 randomly selected evaluation users with the remaining 349 BUCS participants (Table~\ref{tab:paa-representativeness}, Appendix~\ref{app:representativeness}). The NA\_State base rate and mean NA differed with small effect sizes ($r \leq 0.20$); platform mix also differed. The factorial comparisons use the same 50 participants across conditions.

% ---------------------------------------------------------------------------
\subsection{Baselines}
\label{sec:paa-baselines}

\subsubsection{Reference Baselines.}
We report three families of reference baselines to contextualize the aggregate results. First, we include traditional sensing baselines (Random Forest, XGBoost, and Logistic Regression) trained on hourly sensing features across six modalities: motion (stationary, walking, running, automotive minutes), screen (total minutes, session count), GPS (travel distance, time at home, location variance), keyboard (session count, character count), app usage (total and per-category minutes), and light (ambient lux). For each EMA timestamp, features are extracted from hourly windows preceding the prediction point, yielding a flat feature vector (e.g., \texttt{h0\_screen\_on\_min}, \texttt{h1\_motion\_walking\_min}, etc.). The lookback horizon (1, 2, 4, 6, 12, or 24 hours), number of selected features, and model hyperparameters are optimized within the 349-user training set. Second, we include text-only baselines: TF-IDF+Logistic Regression, Bag-of-Words+Logistic Regression, and MiniLM sentence embeddings+Logistic Regression. Third, we include combined baselines (Random Forest and Logistic Regression) over the concatenation of hourly sensing features and diary embeddings. All trained baselines are evaluated on the same held-out 50-user set used for the LLM systems.

These baselines provide context but differ from \pulse{} in training population, feature representation, and data access. The primary comparison therefore remains the factorial contrast between structured and agentic systems evaluated on the same 50 participants under the same current-input conditions.

\subsubsection{LLM Baseline.}
\postreviewedit{CALLM~\cite{wang2026callm} serves as the most directly comparable published diary-based LLM baseline for the same cancer-survivor population and targets. It uses diary text with TF--IDF cross-user retrieval and longitudinal memory.}

% ---------------------------------------------------------------------------
\subsection{Metrics}

Balanced Accuracy (BA) is the primary metric for binary targets, defined as the arithmetic mean of sensitivity and specificity. BA is preferred over standard accuracy for health data with class imbalance, as it gives equal weight to both classes regardless of base rates.

We report pooled, entry-level balanced accuracy across the 50 users, with 95\% bootstrap intervals from 10,000 entry resamples. The reported $p$-values likewise come from entry-level resampling of the BA difference. \reviewrevision{Because this resampling does not account for entries nested within users, we treat the tests as descriptive. Section~\ref{sec:paa-stats} reports per-user dispersion for Auto-Multi.}

% ---------------------------------------------------------------------------
\subsection{Implementation}

All LLM inference uses Claude Sonnet (Anthropic), a transformer-based~\cite{vaswani2017attention} LLM, via the Claude CLI, with up to 16 tool-use turns per prediction. The tools are served via an MCP server launched as a subprocess for each prediction, ensuring per-user data isolation. Inference time ranges from 30 to 90 seconds per prediction depending on the number of tool calls. Each of the five systems (four factorial plus CALLM) produces 4,112 predictions on the 50-user sample. \reviewrevision{All experiments use Claude Sonnet~4.6 (\texttt{claude-sonnet-4-6}). All code, prompts, and tool implementations will be released publicly upon publication.}

% ===========================================================================
% 5. RESULTS
% ===========================================================================
\section{Results}
\label{sec:paa-results}

% ---------------------------------------------------------------------------
\subsection[The 2x2 System Comparison]{The 2$\times$2 System Comparison}
\label{sec:paa-results-factorial}

Within the 2$\times$2 design, the agentic conditions score higher than their structured counterparts in both concurrent-input conditions. \reviewrevision{These are differences between the evaluated conditions: the agentic arms add interactive tool access, personal-baseline comparisons, and selected temporal detail to context shared with the structured arms.} Diary effects are target-dependent and are stated per target below (Sections~\ref{sec:paa-results-intervention}--\ref{sec:paa-diary}). Table~\ref{tab:paa-factorial-results} presents results for the four focus targets and their mean.

\begin{table}[htbp]
\centering
\caption{2$\times$2 system comparison. Balanced accuracy for the four focus targets and their mean; $\Delta$ is the observed difference between the agentic and structured conditions.}
\label{tab:paa-factorial-results}
\small
\resizebox{\textwidth}{!}{%
\begin{tabular}{llccccc}
\toprule
\textbf{Modality} & \textbf{Architecture} & \textbf{PA\_State} & \textbf{NA\_State} & \textbf{ER\_desire} & \textbf{INT\_availability} & \textbf{Mean BA} \\
\midrule
\multirow{3}{*}{No current diary} & Struct-Sense & .502 & .510 & .507 & .542 & .515 \\
 & Auto-Sense & .595 & .591 & .651 & .713 & .637 \\
 & $\Delta$ & +.093 & +.082 & +.145 & +.170 & +.122 \\
\midrule
\multirow{3}{*}{Multimodal} & Struct-Multi & .530 & .666 & .655 & .552 & .601 \\
 & Auto-Multi & \textbf{.720} & \textbf{.716} & \textbf{.743} & \textbf{.722} & \textbf{.725} \\
 & $\Delta$ & +.190 & +.051 & +.088 & +.171 & +.124 \\
\bottomrule
\end{tabular}}%
\end{table}

The analysis reports mean-BA differences of $+0.124$ in the multimodal condition and $+0.122$ in the no-current-diary condition, with entry-level bootstrap $p<.0001$ for each target. \postreviewedit{The largest observed difference is on PA\_State in the multimodal condition ($\Delta = +0.190$). Intervention availability differs by $+0.171$ in the multimodal condition and $+0.170$ in the no-current-diary condition.}

\reviewrevision{\subsubsection{Matched-Call Sensitivity Analysis.}
\label{sec:paa-matched-compute}
Each agentic prediction uses multiple LLM calls (mean 6.2), compared with one call for the structured pipeline. We therefore reran Struct-Multi $k$ times per entry and combined its predictions by majority vote---self-consistency sampling~\cite{wang2023self_consistency}, the standard method for converting additional inference calls on a fixed prompt into accuracy---sweeping $k$ from 1 to 7, where $k{=}7$ approximates the agentic condition's mean call count. The observed mean gap changes from $+.131$ BA at $k{=}1$ to $+.126$ at $k{=}7$ (Appendix~\ref{app:matched-compute}). Repeated structured sampling does not close the gap in this test, although the control matches call count rather than token use or the information available through tools.}

% ---------------------------------------------------------------------------
\subsection{Intervention Opportunity Detection}
\label{sec:paa-results-intervention}

The central empirical question is whether the system can estimate both components of the study's candidate \emph{intervention opportunity}: emotion-regulation desire and self-reported availability. Table~\ref{tab:paa-aggregate} summarizes the four focus targets across systems.

\subsubsection{Regulation Desire (ER\_desire).}
Auto-Multi achieves 0.743 BA, compared with 0.655 for Struct-Multi, 0.629 for CALLM, and 0.651 for Auto-Sense. Thus, the structured--agentic difference within the multimodal condition is 0.088, while the Auto-Multi--Auto-Sense difference associated with providing the current diary is 0.092. Diary text can express the user's current feelings and regulation concerns directly, making the latter pattern plausible, although the comparison does not isolate which part of the diary representation carries the signal.

\subsubsection{Intervention Availability (INT\_availability).}
The corresponding current-diary difference for intervention availability is only 0.010 (Auto-Sense 0.713; Auto-Multi 0.722), compared with 0.092 for regulation desire. Auto-Sense also exceeds CALLM (0.545) and Struct-Multi (0.552) on this target. This target-dependent pattern is notable because the diary contributes much less to stated availability than to regulation desire. \reviewrevision{The simple baselines yield BA .485 from time-of-day features and .540 after adding pre-EMA-hour motion-state minutes (Appendix~\ref{app:baselines}).} Time and recent motion therefore provide a useful reference but do not reproduce the Auto-Sense result.

\subsubsection{Joint Intervention Opportunity Framing.}
Under our study-specific framing, both desire and availability would be relevant at a decision point. Auto-Multi achieves BA values of 0.743 and 0.722 on these two components, respectively. We do not introduce a separate conjunction metric, and the evaluation does not test an intervention-delivery rule. Whether these predictions improve a JITAI decision therefore remains a prospective question.

% ---------------------------------------------------------------------------
\subsection{Aggregate Performance Across All Conditions}

Table~\ref{tab:paa-aggregate} brings the target-level results into one view. It includes the four factorial conditions, the diary-based CALLM baseline, and the conventional reference models, with mean BA shown as a compact summary across the four focus targets.

Struct-Sense is near chance on the four focus targets, whereas Auto-Sense reaches a mean BA of 0.637. In the multimodal condition, mean BA is 0.601 for Struct-Multi and 0.725 for Auto-Multi. The fact that the structured LLM conditions do not automatically exceed the conventional references suggests that replacing a supervised model with a single prompted LLM call is not sufficient by itself.

Auto-Multi has the highest mean BA among the five main systems and is above each reference baseline in the aggregate table. These broader comparisons provide context but are not directly controlled: the systems differ in training population, feature access, temporal resolution, and retrieval. The within-design contrasts therefore remain the more informative comparisons.

\reviewrevision{MiniLM is the strongest lightweight reference. Auto-Multi exceeds MiniLM+LR by $.027$ BA on PA, $.037$ on NA, $.062$ on regulation desire, and $.119$ on intervention availability. MiniLM+LR is trained on roughly 16,000 labeled entries from the other 349 participants, whereas Auto-Multi receives outcome-labeled examples through retrieval at inference time. The comparison is therefore descriptive rather than directly controlled (Appendix~\ref{app:baselines}).}

% ---------------------------------------------------------------------------
\subsection{Affect State Detection}
\label{sec:paa-results-affect}

Affect state detection follows the same factorial pattern (Table~\ref{tab:paa-factorial-results}). PA\_State differs by 0.190 BA in the multimodal condition (Auto-Multi: 0.720; Struct-Multi: 0.530), the largest single-target difference in the design. \reviewrevision{Struct-Multi predicts the positive class on 7\% of entries against a roughly 55\% base rate (sensitivity .09, specificity .95), while Auto-Multi predicts it on 54\%. One plausible explanation is that personal-baseline and retrieved-outcome cues help the agentic condition recognize positive states more often, although these channels were not independently manipulated.} The result is therefore driven not merely by a small improvement around the decision boundary, but by a large difference in how often the positive state is recognized.

For NA\_State, the BA values are 0.641 for CALLM, 0.666 for Struct-Multi, and 0.716 for Auto-Multi; the Struct-Multi--Auto-Multi difference is 0.051. Diary text is often directly expressive of negative emotion, which may help explain why the structured and diary-only systems are stronger here than on PA\_State. Auto-Multi nevertheless remains higher, indicating that the additional context available in that condition contributes information beyond the diary-only result.

% ---------------------------------------------------------------------------
\subsection{No Current Diary vs.\ Multimodal: Observed Difference}
\label{sec:paa-diary}

Table~\ref{tab:paa-diary} reports the observed difference associated with including the current diary within the agentic condition. Mean BA changes from 0.637 to 0.725. The differences are 0.125 for both affect states and 0.092 for regulation desire, but only 0.010 for intervention availability. The current diary is therefore associated with a substantial change for affect and regulation desire, whereas the availability result is comparatively stable when the current diary is withheld. This distinction motivates the construct-level discussion in Section~\ref{sec:paa-discussion} rather than a single conclusion about the value of diary input.

% ---------------------------------------------------------------------------
\subsection{Statistical Significance and Effect Sizes}
\label{sec:paa-stats}

The entry-level bootstrap analysis uses 10,000 resamples and reports $p<.0001$ for all four target contrasts in both concurrent-input conditions. \reviewrevision{Table~\ref{tab:paa-statistics} reports the observed differences and 95\% bootstrap intervals. The directional tests evaluate whether BA is higher for the agentic condition than for the structured condition. Because entry resampling ignores within-user dependence, we treat these tests as descriptive.}

\begin{table}[t]
\centering
\caption{Descriptive pairwise contrasts. Bootstrap tests resample entries (10,000 resamples) and do not account for within-user dependence.}
\label{tab:paa-statistics}
\small
\begin{tabular}{llccc}
\toprule
\textbf{Comparison} & \textbf{Target} & $\Delta$ \textbf{BA} & \textbf{$p$-value} & \textbf{95\% CI} \\
\midrule
\multirow{5}{*}{Auto-Multi vs.\ Struct-Multi} & PA\_State & +.190 & $<.0001$ & [+.173, +.208] \\
 & NA\_State & +.051 & $<.0001$ & [+.037, +.064] \\
 & ER\_desire & +.088 & $<.0001$ & [+.072, +.105] \\
 & INT\_availability & +.171 & $<.0001$ & [+.152, +.190] \\
 & Mean BA & +.124 & $<.0001$ & [+.113, +.136] \\
\midrule
\multirow{5}{*}{Auto-Sense vs.\ Struct-Sense} & PA\_State & +.093 & $<.0001$ & [+.077, +.110] \\
 & NA\_State & +.082 & $<.0001$ & [+.064, +.099] \\
 & ER\_desire & +.145 & $<.0001$ & [+.128, +.161] \\
 & INT\_availability & +.170 & $<.0001$ & [+.153, +.186] \\
 & Mean BA & +.122 & $<.0001$ & [+.112, +.133] \\
\bottomrule
\end{tabular}
\end{table}

\reviewrevision{Table~\ref{tab:paa-user-dispersion} complements the pooled contrasts with across-user dispersion for Auto-Multi; SDs range from .091 to .116 across the four focus targets.}

% ---------------------------------------------------------------------------
\subsection{Continuous-Score Bias and Failure Modes}

\postreviewedit{The continuous outputs show systematic bias. Auto-Multi over-predicts Negative Affect (predicted mean ${\sim}$6.2 vs.\ observed mean ${\sim}$3.2; MAE 4.21) and under-predicts Positive Affect (predicted mean ${\sim}$16.7 vs.\ observed mean ${\sim}$18.0; MAE 5.42), with compressed PA variance (predicted SD 4.7 vs.\ observed SD 8.1). ER desire has MAE 1.74 and correlation $r=.61$; these are agreement measures, not a calibration test.}

\postreviewedit{The model emits continuous scores and binary state labels as separate fields; the binary results therefore do not establish that the continuous scores are well calibrated. The non-agentic LLM systems show smaller continuous-scale error on some targets but lower binary BA. We therefore report the continuous and binary results separately. The continuous negativity bias remains a failure mode.}

% ---------------------------------------------------------------------------
\subsection{Agentic Investigation Patterns}
\label{sec:paa-traces}

\subsubsection{Tool Usage Patterns.} Three core tools appear in 99.7\% of Auto-Multi predictions. They are \path{get_daily_summary}, \path{get_behavioral_timeline}, and \path{find_peer_cases}; the last follows a standing prompt instruction. \path{compare_to_baseline} appears in 60.7\%, and the remaining tools in 4.3--14.2\%.

\reviewrevision{Two analyses examine negativity bias. The system predicts elevated NA on 30.0\% of entries against an observed base rate of roughly 32\%, with a 16.9\% false-positive rate. Tool--prediction co-occurrence is higher for several optional drill-down tools, but this correlational pattern does not show whether tool use causes negative predictions or follows initial suspicion. In a controlled neutral-prompt comparison, the observed negative-prediction rate, false-positive rate, and balanced accuracy each change by less than .01 (Appendix~\ref{app:negbias}).}

\subsubsection{Confidence Stratification.} Each prediction includes a self-reported confidence score between 0 (not at all confident) and 1 (entirely confident) (mean 0.76, SD = 0.10 sample-wise). Accuracy differs across confidence bands: predictions below 0.65 achieve 66.9\% average accuracy across the four focus targets, predictions from 0.65--0.80 achieve 74.1\%, and predictions above 0.80 achieve 82.5\%. These bands describe an association, not probabilistic calibration. We examine regulation desire in Section~\ref{sec:paa-confidence-over-time}.

\subsubsection{Computational Footprint.} Each prediction consumes approximately 88,000 input tokens and 4,000 output tokens, with session memory growing to roughly 50K characters per user by the end of the evaluation. Each prediction takes 30--90 seconds wall-clock time, including MCP server startup, multi-turn tool use, and response generation.

\reviewrevision{\subsubsection{Session-Memory Reflection Ablation.}
\label{sec:paa-memory-contribution}
Removing the per-entry reflection document while retaining the other memory and history channels reduces regulation-desire BA by approximately $.025$; no consistent difference appears for PA, NA, or intervention availability (Appendix~\ref{app:memory}). This analysis isolates the reflection document rather than the full memory system.}

\postreviewedit{Appendix~\ref{app:cases} presents one constructed interface example and two selected rationales that illustrate which historical-outcome channels appeared in model context.}

% ---------------------------------------------------------------------------
\subsection{Confidence over Entry Position}
\label{sec:paa-tool-evolution}
\label{sec:paa-confidence-over-time}

\postreviewedit{Self-reported confidence rises from 0.69 to 0.76 across the first 70 entry positions. Because entry position also tracks accumulated history and sensing coverage, this trend is descriptive.}

\begin{figure}[t]
\centering
\begin{subfigure}{0.48\textwidth}
    \centering
    \includegraphics[width=\linewidth]{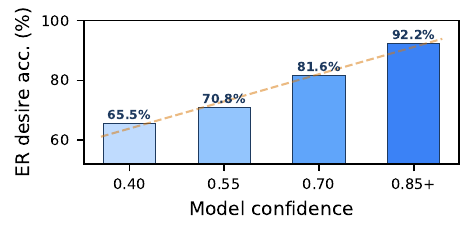}
    \caption{Confidence stratification.}\label{fig:confidence-calibration}
\end{subfigure}\hfill
\begin{subfigure}{0.48\textwidth}
    \centering
    \includegraphics[width=\linewidth]{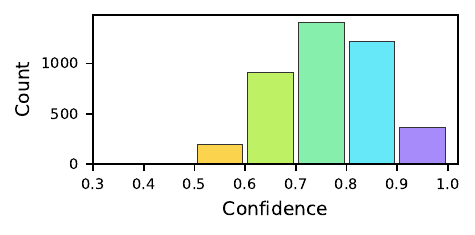}
    \caption{Confidence distribution.}\label{fig:confidence-distribution}
\end{subfigure}
\caption{Self-reported confidence. (a)~Regulation-desire balanced accuracy by confidence band, increasing from 65.5\% to 92.2\%; this is stratification rather than a probabilistic calibration test. (b)~Confidence distribution.}
\Description{Two panels summarize self-reported confidence: balanced accuracy for ER desire increases across four confidence bands, and a histogram shows the distribution of confidence scores.}
\label{fig:agent-behavior}
\end{figure}

Confidence stratifies accuracy (Figure~\ref{fig:agent-behavior}a): for regulation desire, balanced accuracy rises across confidence bands, from 65.5\% (confidence ${\approx}$0.40, $n{=}55$) through 70.8\% and 81.6\% to 92.2\% at confidence ${\geq}$0.85 ($n{=}348$). Verbalized confidence is known to be informative but often miscalibrated~\cite{xiong2024confidence}, so we test only whether accuracy rises with reported confidence.

% ===========================================================================
% 6. DISCUSSION
% ===========================================================================
\section{Discussion}
\label{sec:paa-discussion}

% ---------------------------------------------------------------------------
\reviewrevision{\subsection{Interpreting the Structured--Agentic Difference}}

\reviewrevision{The structured--agentic comparison changes both interaction style and available information. The structured condition receives a preassembled sensing summary, a longitudinal memory document, and outcome-labeled peer examples in a single prompt. The agentic condition instead works through a multi-turn interface: it can request additional temporal detail, compare an observation with a personal baseline, and retrieve similar days before producing a prediction. We therefore interpret the result as a comparison between the two implemented systems.}

One candidate explanation is \emph{selective evidence assembly}. A fixed summary presents the same categories in the same order for every entry. The interactive condition can begin with a broad overview and then spend additional queries on the modalities or time windows that appear relevant for that person and moment. A possible sleep disruption, unusual screen pattern, or mobility change can lead to different follow-up queries. This flexibility does not guarantee that the selected evidence is useful, but it allows the available context to depend on the entry rather than solely on a template.

A second candidate explanation is explicit access to \emph{personal comparison and episodic examples}. The \path{compare_to_baseline} tool turns an absolute observation into a within-person deviation, while \path{find_similar_days} returns earlier same-user cases and their outcomes. The agentic condition can decide when to request these comparisons and combine them with outcome-labeled peer cases. The structured conditions also receive peer examples, so peer retrieval alone cannot account for the observed difference; the distinction lies in the broader combination of interactive querying, personal comparison, and accessible history.

These design choices instantiate contextual behavioral signal interpretation~\cite{narayanan2013bsp, xu2024penetrative_ai}: meaning is assembled from a person's recent behavior, baseline, and related episodes rather than from a single population-level feature vector. \reviewrevision{Because these components were not independently ablated, they remain candidate explanations. The matched-call analysis (Section~\ref{sec:paa-matched-compute}) shows that repeated structured sampling does not close the observed gap.}

% ---------------------------------------------------------------------------
\subsection{The Diary Paradox and Graceful Degradation}

Prompt-level EMA work shows that affective states can predict subsequent nonresponse in some cohorts~\cite{murray2023ema}, and broader mHealth work documents bias from non-random missingness~\cite{goldberg2021missing}. We use ``diary paradox'' to describe the resulting design tension: diary text can be highly informative when it is present, yet a system that depends on the current diary cannot use that text when a prompt is omitted. \reviewrevision{The present benchmark evaluates the first technical step in that problem---withholding the current diary on completed EMA occasions---rather than performance at the omitted occasions themselves.}

With the current diary withheld, mean BA drops from 0.725 to 0.637. The size of the difference depends strongly on the target. It is 0.125 for both affect states and 0.092 for regulation desire, but only 0.010 for intervention availability. The no-current-diary condition therefore preserves much more of the availability result than of the affect and desire results. This pattern is more informative than the aggregate difference because it shows where a current diary changes the prediction problem most.

This result supports a fallback design in which the system can still return an estimate when the current diary is unavailable, while recognizing that the estimate may degrade differently across targets. It does not establish that such a fallback is adequate for intervention delivery. The same circumstances that reduce diary completion may also change sensing coverage: a phone may be left behind, turned off, or used differently. A prospective study should therefore include prompted occasions whether or not a diary was completed, record the sensing available at those moments, and evaluate whether target-specific performance changes under actual nonresponse.

% ---------------------------------------------------------------------------
\subsection{Regulation Desire and Stated Availability}

The observed current-diary difference is larger for regulation desire ($+0.092$) than for intervention availability ($+0.010$), suggesting that the two labels capture different predictive information. Regulation desire records whether the participant wants to regulate their current emotional state. Intervention availability records whether the participant reports being available to engage with support. They are related at a decision point, but neither is a substitute for the other: someone may want to change how they feel while being occupied, or may be available without currently wanting to regulate emotion.

One plausible interpretation of the result is that diary text is closer to the internal experience represented by regulation desire. A diary can directly express distress, conflict, gratitude, or uncertainty, all of which may inform a desire to regulate emotion. Availability may align more strongly with contextual patterns such as time, activity, and recent phone interaction, which remain visible without the current diary. \reviewrevision{The simple-baseline analysis, however, shows that time of day and recent motion alone reach only .485 and .540 BA, respectively; these two cues do not reproduce the availability result.}

Taken together, these results support modeling and evaluating desire and stated availability separately as candidate tailoring variables~\cite{nahum2018jitai, kunzler2019receptivity, mishra2021receptivity}. A single receptivity label or aggregate score would obscure their different associations with current-diary input.

% ---------------------------------------------------------------------------
\subsection{Outcome-Labeled Peer Retrieval}

The cross-user RAG mechanism differs from factual RAG in question answering~\cite{lewis2020retrieval, gao2024ragsurvey}. Rather than retrieving documents that contain an answer, it returns cases selected as similar together with their observed study outcomes. We therefore call it \emph{outcome-labeled peer retrieval}. These examples provide a local empirical reference: the model can see how affect, desire, and availability were labeled for participants whose diary text or sensing fingerprint resembles the current case.

This design is potentially useful when the same behavioral measurement has different meanings across contexts. Retrieved outcomes can supply cohort-specific examples alongside the model's general knowledge and the person's own history. In the agentic condition, peer retrieval can also be considered together with same-user similar days and explicit baseline comparisons. That combination offers one route for relating a current pattern to both idiographic and population context.

\reviewrevision{Because peer retrieval exposes the model to outcome distributions from the cancer-survivor cohort, its utility may depend on how well the reference cohort matches the target population. A database dominated by one demographic group or care context may provide less relevant analogies for another. Peer retrieval is part of the evaluated system and was not independently ablated, so we do not attribute the performance difference to it alone.}

% ---------------------------------------------------------------------------
\subsection{Limitations}

\textbf{Retrospective evaluation.} \pulse{} was evaluated on existing BUCS data, not in a prospective deployment. Real-time use would introduce latency, network, model-availability, and interface constraints that are not captured here and require prospective evaluation.

\textbf{Sample size and dependence.} The evaluation includes 50 users. \reviewrevision{Because entries are repeated within users, the pooled entry bootstrap does not support population-level significance claims. The sample is also predominantly female (90\%) and White (94\%), which limits generalizability.}

\textbf{Baseline and condition comparability.} The broader baselines differ in training data and feature access. \reviewrevision{Within the factorial design, the structured and agentic conditions share users and current-input modality but not identical information interfaces. The design therefore supports system comparison rather than a causal claim about reasoning architecture.}

\textbf{Model dependency.} All experiments used a single LLM (Claude Sonnet). While another tool-capable model could instantiate the system, performance may vary. Open-weight alternatives such as LLaMA~\cite{touvron2023llama} could improve reproducibility and support research on local deployment. The MCP tool interface facilitates future cross-model evaluation.

\textbf{Inference cost.} Each prediction requires 30--90 seconds and roughly 90K tokens of LLM inference, making full-cohort evaluation expensive and real-time deployment cost-sensitive. This cost constrained our evaluation to 50 of 399 available users. Model distillation or on-device inference could reduce cost, but whether they preserve the observed performance is unknown.

% ---------------------------------------------------------------------------
\subsection{Implications for JITAI Design}

The study suggests four priorities for JITAI research~\cite{nahum2018jitai}:

\begin{enumerate}
    \item \textbf{Model desire and availability separately.} A decision point may depend on both the desire to regulate emotion and the practical availability to engage. Their different associations with the current diary argue against collapsing them into a single receptivity target before each component is understood.

    \item \reviewrevision{\textbf{Compare interactive and preassembled information access.} The agentic interface lets the model request personal comparisons and follow up on selected observations. Future studies should compare it with non-agentic systems that receive comparable information so that the value of interaction, retrieval, and additional context can be distinguished.}

    \item \textbf{Plan explicitly for missing current input.} A system may use the current diary when available while retaining a no-current-diary pathway. The target-dependent differences here show why the quality of that fallback should be evaluated separately for affect, regulation desire, and availability rather than summarized by one average.

    \item \reviewrevision{\textbf{Evaluate outcome-labeled retrieval as a component.} Similar cases can provide useful empirical context, but they also carry cohort-specific outcome distributions. Retrieval databases, similarity functions, and cross-population transfer should therefore be examined directly before they inform intervention timing.}
\end{enumerate}

These priorities complement rather than replace established JITAI development methods. Observational studies can identify candidate tailoring variables, and microrandomized trials can estimate the proximal effects of intervention options at decision points~\cite{klasnja2015mrt}. \pulse{} addresses a different layer: how behavioral observations can be assembled into state estimates when inputs vary across moments. The resulting decision rules and interventions would still require prospective experimental evaluation.

% ---------------------------------------------------------------------------
\subsection{Ethical Considerations}

This study was conducted under IRB oversight, and all participants provided informed consent for passive smartphone sensing and EMA data collection. Data were de-identified before analysis, with participant IDs replaced by numeric study identifiers.

The combination of passive sensing data and LLM-based inference raises privacy considerations. Sensing data, although it contains no message or diary content, can reveal sensitive behavioral patterns (location history, communication patterns, sleep disruption). As Coravos et al.~\cite{coravos2019digital} emphasize, digital biomarkers must be developed with safety and effectiveness frameworks from the outset. In this study, data were processed in a research context with appropriate safeguards. Future deployment would require careful attention to data handling practices~\cite{privacy_fl_health2024}, user control over data sharing, transparency about how predictions are generated~\cite{explainability_llm2025}, and compliance with applicable privacy regulations.

We emphasize that \pulse{} is a prediction layer, not a clinical intervention or diagnostic system~\cite{vaidyam2019chatbots, ai_chatbot_meta2024}. Its predictions are intended to inform intervention delivery decisions within a broader JITAI framework, not to replace clinical assessment. The system should not be the sole determinant of clinical action. False-positive predictions could increase unnecessary check-ins or alerts, so deployment studies should evaluate alert burden alongside missed opportunities for support.

\postreviewedit{The use of LLM APIs for behavioral-health data raises questions about transmission and storage. In the current implementation, model inputs are de-identified. On-device models~\cite{wang2024slm, zhang2024ondevice, edge_llm_survey2024} could reduce external transmission and, under a fully local architecture, eliminate model-input transmission to an external provider.}

% ===========================================================================
% 7. FUTURE WORK
% ===========================================================================
\section{Future Work}
\label{sec:paa-future}

Several directions follow from this work. First, a prospective study should evaluate the system at actual decision points, including occasions with no completed diary, and use a microrandomized or related design if predictions are connected to intervention delivery~\cite{klasnja2015mrt}. Such a study should record sensor coverage and retain outcomes independently of whether the participant answers the current prompt. Second, a future component study could provide comparable information and inference budgets while varying personal-baseline queries, same-user retrieval, peer retrieval, and the reflection document separately. This would test the candidate explanations in Section~\ref{sec:paa-discussion} beyond the full-system comparison reported here. Third, cross-model and local-model experiments can test whether the observed patterns depend on Claude Sonnet and whether smaller or on-device models can support the same tool interface~\cite{wang2024slm, edge_llm_survey2024, zhang2024ondevice}. Finally, evaluation beyond cancer survivors should examine how the tool design, retrieved cohort, and target definitions transfer to other populations and sensing settings.

% ===========================================================================
% 8. CONCLUSION
% ===========================================================================
\section{Conclusion}
\label{sec:paa-conclusion}

This paper introduces \pulse{}, an LLM system that combines multi-turn sensing queries, personal-baseline comparisons, and outcome-labeled historical retrieval. Rather than relying on one preassembled summary, the model can request additional behavioral context before returning the study-target predictions.

In a 50-user retrospective benchmark, the agentic conditions score above structured single-pass systems under both concurrent-input settings. Because the conditions also differ in accessible information, these results characterize the implemented systems rather than isolate one causal mechanism. The matched-call analysis narrows one alternative explanation, while the remaining contribution of interaction, personal comparison, retrieval, and memory requires component-level study.

The association with the current diary is also target-dependent. Emotion-regulation desire and affect change appreciably when the diary is withheld, whereas self-reported availability changes little in this evaluation. This distinction motivates separate treatment of desire and availability and a fallback pathway when current self-report is absent. More broadly, the study shows how interactive sensing investigation can be operationalized for affect and availability prediction. Its value for JITAI decisions, omitted-prompt settings, and other populations now requires prospective evaluation with comparable information access across conditions.

\bibliographystyle{ACM-Reference-Format}
\bibliography{references}

%% ===========================================================================
%% APPENDICES
%% ===========================================================================
\appendix

\reviewrevision{\section{Prompt Structure}\label{app:prompts}}

\reviewrevision{The complete system prompts, the reflection prompt, and all tool implementations will be released verbatim with the source code upon publication. Here we document the prompt structure; Table~\ref{tab:paa-tools} presents the tool interfaces.}

\reviewrevision{\paragraph{Agentic system prompts.} Each agentic prompt contains, in order: a role definition; the task statement for the concurrent-input condition (multimodal prompts prioritize the current diary, whereas no-current-diary prompts omit it); an investigation-strategy sketch over the eight tools, including a standing instruction to retrieve peer cases; the study's conjunction rule for desire and availability; a description of the EMA-derived composite used by the reflection step; and the structured JSON output schema covering binary targets, three continuous scores, a reasoning trace, and a confidence value between 0 and 1, where 0 means not at all confident and 1 means entirely confident, with no meaning attached to intermediate values, following standard verbalized-confidence elicitation~\cite{lin2022verbalized,xiong2024confidence}. Historical diary-derived outcomes can still enter no-current-diary prompts through memory and retrieval.}

\reviewrevision{\textbf{Neutral-prompt variant.} The variant used in the negativity-bias control (Section~\ref{sec:paa-traces}) differs from the standard multimodal prompt in exactly two passages: ``look for behavioral evidence that confirms or contradicts what the diary suggests \ldots{} Discrepancies are informative'' becomes ``consider cross-modal patterns: how does their behavior relate to what they wrote?'', and ``check if today's behavior is unusual for this person'' becomes ``characterize how today's behavior compares to this person's typical patterns.'' The public release adopts the neutral phrasing; the reported metrics changed by less than .01 in this comparison.}

\reviewrevision{\textbf{Reflection prompt.} After each prediction, a lightweight model call produces a 1--2 sentence lesson for the session memory. The prompt provides the investigation summary, the prediction, and an EMA-derived binary composite of regulation desire and availability; it does not provide the two component labels separately or the affect scores. The accumulated reflection document is passed in full at subsequent entries and grows to roughly 50K characters per user.}

\reviewrevision{We document the prompt structure here. Complete prompts and the agentic system code will be released publicly upon publication.}

\reviewrevision{\section{Supplementary Results and Robustness Analyses}\label{app:robustness}}

This appendix provides the expanded results supporting the main comparisons. Tables~\ref{tab:paa-aggregate}--\ref{tab:paa-user-dispersion} report system-level performance, the observed differences associated with providing the current diary, and across-user dispersion, respectively. The remaining sections present the matched-call control, session-memory ablation, cold-start analysis, negativity-bias control, reference baselines, temporal-boundary implementation, and evaluation-sample representativeness analysis.

\begin{table}[t]
\centering
\caption{Per-target and aggregate performance across systems. Balanced accuracy for the four focus targets and their mean. Reference baselines are included for context; the main comparative claim comes from the within-study 2$\times$2 comparison.}
\label{tab:paa-aggregate}
\small
\resizebox{\textwidth}{!}{%
\begin{tabular}{llccccc}
\toprule
\textbf{System} & \textbf{Input} & \textbf{ER\_desire} & \textbf{INT\_availability} & \textbf{PA\_State} & \textbf{NA\_State} & \textbf{Mean BA} \\
\midrule
\multicolumn{7}{l}{\emph{Reference baselines}} \\
Random Forest & Sensing & .501 & .496 & .520 & .501 & .505 \\
XGBoost & Sensing & .493 & .508 & .525 & .512 & .510 \\
Combined RF & Sensing + diary & .574 & .542 & .666 & .596 & .595 \\
MiniLM & Sensing + diary & .681 & .603 & .693 & .679 & .664 \\
\midrule
\multicolumn{7}{l}{\emph{LLM systems}} \\
CALLM & Diary text & .629 & .545 & .534 & .641 & .587 \\
Struct-Sense & Sensing & .507 & .542 & .502 & .510 & .515 \\
Auto-Sense & Sensing & .651 & .713 & .595 & .591 & .637 \\
Struct-Multi & Sensing + diary & .655 & .552 & .530 & .666 & .601 \\
Auto-Multi & Sensing + diary & .743 & .722 & .720 & .716 & .725 \\
\bottomrule
\end{tabular}}%
\end{table}

\begin{table}[h]
\centering
\caption{Observed difference between multimodal and no-current-diary agentic conditions (Auto-Multi minus Auto-Sense). Both conditions may access historical self-report outcomes.}
\label{tab:paa-diary}
\small
\begin{tabular}{lccc}
\toprule
\textbf{Target} & \textbf{Auto-Sense} & \textbf{Auto-Multi} & $\Delta$ \\
\midrule
PA\_State & .595 & .720 & +.125 \\
NA\_State & .591 & .716 & +.125 \\
ER\_desire & .651 & .743 & +.092 \\
INT\_availability & .713 & .722 & +.010 \\
\midrule
Mean BA & .637 & .725 & +.088 \\
\bottomrule
\end{tabular}
\end{table}

\begin{table}[H]
\centering
% \color{blue}
\caption{\reviewrevision{Auto-Multi per-user balanced accuracy across all 50 evaluation users.}}
\label{tab:paa-user-dispersion}
\small
\begin{tabular}{lccc}
\toprule
\textbf{Target} & \textbf{Mean} & \textbf{SD} & \textbf{$n$ users} \\
\midrule
PA\_State & .682 & .095 & 50 \\
NA\_State & .704 & .093 & 50 \\
ER\_desire & .698 & .091 & 50 \\
INT\_availability & .603 & .116 & 50 \\
\bottomrule
\end{tabular}
\end{table}

\reviewrevision{\subsection{Matched-Call Sensitivity Analysis: Per-Target Results}\label{app:matched-compute}}

\begin{figure}[H]
\centering
\color{blue}
\includegraphics[width=0.6\columnwidth]{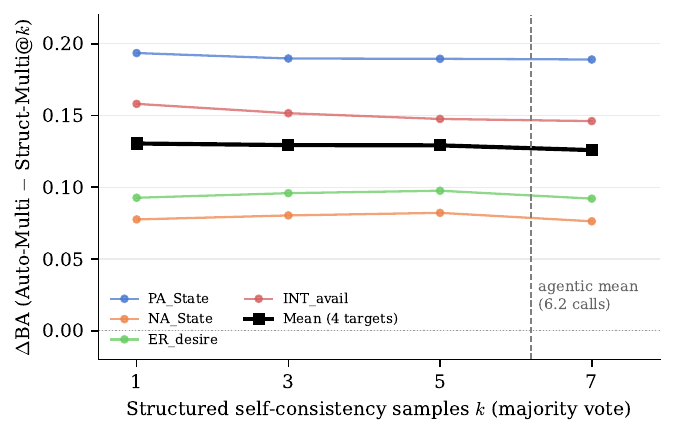}
\caption{\reviewrevision{Matched-call sensitivity analysis. Agentic advantage ($\Delta$BA, Auto-Multi minus Struct-Multi with $k$-sample majority voting) as a function of the structured system's inference budget. The dashed line marks the agentic system's mean number of LLM calls (6.2).}}
\Description{Line chart showing that the agentic balanced-accuracy advantage remains nearly constant as structured self-consistency increases from one to seven samples.}
\label{fig:kcurve}
\end{figure}

\reviewrevision{All control analyses use Claude Sonnet~4.6. Table~\ref{tab:app-kcurve} gives the matched-call results per target, and Figure~\ref{fig:kcurve} shows how the observed difference changes as $k$ increases. Both arms are independent runs; per-target values differ slightly from Table~\ref{tab:paa-factorial-results} under nondeterministic decoding. The analysis matches call count only; it does not match tokens, latency, or the information available through tools.}

\begin{table}[h]
\centering
% \color{blue}
\caption{\reviewrevision{Matched-call sensitivity analysis: observed difference ($\Delta$BA, Auto-Multi minus Struct-Multi with $k$-sample majority voting).}}
\label{tab:app-kcurve}
\small
\begin{tabular}{lcccc}
\toprule
\textbf{Target} & $k{=}1$ & $k{=}3$ & $k{=}5$ & $k{=}7$ \\
\midrule
PA\_State & $+.194$ & $+.190$ & $+.190$ & $+.189$ \\
NA\_State & $+.078$ & $+.080$ & $+.082$ & $+.076$ \\
ER\_desire & $+.093$ & $+.096$ & $+.098$ & $+.092$ \\
INT\_availability & $+.158$ & $+.152$ & $+.148$ & $+.146$ \\
\midrule
\textbf{Mean} & $+.131$ & $+.129$ & $+.129$ & $+.126$ \\
\bottomrule
\end{tabular}
\end{table}

\reviewrevision{\subsection{Session Memory: Per-Target Contribution and Accumulation}\label{app:memory}}

\begin{figure}[H]
\centering
\color{blue}
\includegraphics[width=0.55\columnwidth]{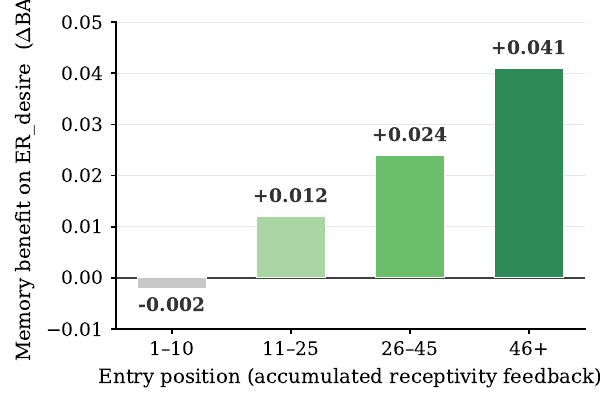}
\caption{\reviewrevision{ER\_desire difference ($\Delta$BA, with reflection document minus without) by entry position. The difference grows from near zero in the first ten entries to $+.041$ beyond entry 45.}}
\Description{Bar chart showing the balanced-accuracy difference for ER desire across four entry-position groups; the benefit of the reflection document grows with later entry position.}
\label{fig:memory-dose}
\end{figure}

\reviewrevision{Removing the per-entry reflection document while retaining the initial memory and historical-outcome tools reduces ER\_desire BA by approximately $.025$. The other targets do not show a consistent pattern. Figure~\ref{fig:memory-dose} shows how the ER\_desire difference varies with entry position.}

\reviewrevision{\subsection{Cold-Start Performance}\label{app:coldstart}}

\reviewrevision{Relative to entries 41+, pooled four-target BA is $-.029$ over entries 1--10 and $-.020$ over entries 11--20. A similar difference appears with the reflection document ablated ($-.043$ in the earliest bin). Entry position also tracks accumulated history and sensing coverage.}

\reviewrevision{\subsection{Negativity-Bias Analyses}\label{app:negbias}}

\reviewrevision{Across all evaluation investigations, the system predicts elevated NA on 30.0\% of entries (ground-truth base rate ${\sim}$32\%), with an NA false-positive rate of 16.9\%. Universal tools show no elevated co-occurrence with negative predictions (lift 1.0), whereas several optional drill-down tools co-occur more often (\texttt{query\_raw\_events} 2.0, \texttt{query\_sensing} 1.6, \texttt{get\_receptivity\_history} 1.4). This pattern may reflect follow-up after an initial negative cue; co-occurrence alone does not establish direction. In the neutral-prompt condition (Appendix~\ref{app:prompts}), the negative-prediction rate (.304 vs.\ .300), false-positive rate (.170 vs.\ .169), and balanced accuracy (${\Delta}<.01$) remain nearly unchanged.}

\reviewrevision{\subsection{Simple Availability Baseline and Lightweight-Model Comparison}\label{app:baselines}}

\reviewrevision{The simple availability baselines use logistic regression trained on the remaining 349 participants and evaluated on the same 50-user set: time-of-day features alone (sine/cosine hour, weekend) give BA .485, while adding pre-EMA-hour motion-state minutes gives BA .540. Pre-EMA motion coverage is missing for 21\% of entries and is handled with a missingness flag. MiniLM+LR, trained on roughly 16,000 labeled entries from the same 349 participants, trails Auto-Multi by $+.027$ (PA), $+.037$ (NA), $+.062$ (ER\_desire), and $+.119$ (INT\_availability).}

\reviewrevision{\subsection{Temporal-Boundary Enforcement}\label{app:audit}}

\reviewrevision{The current EMA timestamp is a hard cutoff in deterministic data-access code: sensing records, historical outcomes, memory, retrieval results, and structured summaries are restricted to information available before that cutoff.}

\subsection{Representativeness of the Evaluation Sample}
\label{app:representativeness}

Table~\ref{tab:paa-representativeness} compares the 50-user evaluation sample with the remaining 349 eligible BUCS participants. The NA\_State base rate and mean NA differ with small effect sizes; platform mix also differs.

\begin{table}[h]
\centering
\caption{Representativeness of the 50 test users versus the remaining 349 BUCS participants. Mann-Whitney U tests for continuous measures; chi-square for categorical. Effect sizes reported as $r$ (rank-biserial correlation).}
\label{tab:paa-representativeness}
\small
\begin{tabular}{lcccc}
\toprule
\textbf{Measure} & \textbf{Test} & \textbf{Rest} & \textbf{$p$} & \textbf{$r$} \\
 & ($N{=}50$) & ($N{=}349$) & & \\
\midrule
PA\_State base rate & 0.545 & 0.561 & .101 & 0.14 \\
NA\_State base rate & 0.317 & 0.362 & .028* & 0.19 \\
ER\_desire base rate & 0.350 & 0.385 & .256 & 0.10 \\
INT\_availability base rate & 0.643 & 0.593 & .295 & 0.09 \\
Mean PA & 17.86 & 17.27 & .527 & 0.06 \\
Mean NA & 3.23 & 4.38 & .024* & 0.20 \\
Mean ER\_desire & 2.32 & 2.71 & .255 & 0.10 \\
Platform (\% Android) & 36\% & 25\% & .027* & -- \\
CV group distribution & -- & -- & .292 & -- \\
\bottomrule
\multicolumn{5}{l}{\small *$p < .05$. Effect sizes for significant differences are small ($r \leq 0.20$).}
\end{tabular}
\end{table}

\section{Case Studies}
\label{app:cases}

\subsection{Example: Structured vs.\ Agentic on the Same Entry.}
\postreviewedit{The following constructed example illustrates the interface difference; it is not an evaluation trace. Consider a diary entry reading: ``Feeling tired but trying to stay positive after my appointment.'' A structured agent would receive a pre-formatted daily summary, while an agentic agent could request additional baseline and retrieval results.}

An illustrative agentic sequence is:
\begin{enumerate}
    \item Reads the diary and notes the tension between tiredness and positive coping.
    \item Calls \texttt{get\_daily\_summary} and observes the day's overall patterns.
    \item Notices low walking minutes and calls \texttt{compare\_to\_baseline} on walking, discovering it is 1.8 standard deviations below this user's personal mean.
    \item Calls \texttt{get\_behavioral\_timeline} and observes a gap in activity during the morning hours, consistent with a medical appointment.
    \item Calls \texttt{query\_sensing} on screen usage and finds a prolonged phone session during the morning gap, consistent with time spent waiting.
    \item Calls \texttt{find\_similar\_days}; in this illustration, suppose it retrieves three past days with low mobility and medical context, when the user's affect was lower than the diary language alone suggested.
    \item Calls \texttt{find\_peer\_cases} with the diary text; suppose the returned cases pair similar ``trying to stay positive'' language with moderate-to-low positive affect.
    \item Synthesizes the potentially conflicting cues: the diary conveys positive coping, while the behavioral and retrieved context suggests that positive affect may still be below the user's personal mean.
\end{enumerate}

\postreviewedit{This example shows how interactive queries could expose explicit baseline and retrieval anchors. The structured condition receives some related context in preassembled form but cannot choose or iteratively refine queries. Because this is a constructed illustration, it is not evidence that these steps caused the observed performance difference.}

\subsection{Selected Rationale: Personal and Peer Outcome History.}

\label{sec:paa-case-funeral}

\postreviewedit{The following trace shows a prediction in which the model's stated rationale places personal history above peer outcomes. It illustrates how the model uses the available context rather than isolating the contribution of session memory.}

\postreviewedit{Participant~089 had a positive EMA-derived desire-and-availability composite on 6\% of 39 earlier entries. On one December entry, the diary reads: \emph{``Funeral of close friend today. Worried about mother's declining health.''} The following five steps summarize the model's rationale, not verified psychological facts:}

\begin{enumerate}
    \item \textbf{Diary analysis.} The agent identifies two compounding stressors---grief and caregiving worry---and forms an initial hypothesis: elevated negative affect and likely elevated ER~desire.
    \item \postreviewedit{\textbf{Behavioral overview.} \path{get_behavioral_timeline} and \path{get_daily_summary} return a stationary interval and 2~minutes of screen use by mid-morning. The model interprets these values as possible withdrawal.}
    \item \postreviewedit{\textbf{Baseline comparison.} \path{compare_to_baseline} reports 2~minutes of screen use against a typical value of roughly 14~minutes per hour. The model labels this as disengagement; the tool itself only returns the comparison.}
    \item \postreviewedit{\textbf{Peer outcomes.} \path{find_peer_cases} with ``funeral friend'' returns grief-related diary cases, many with elevated ER~desire labels. The model treats these labels as evidence favoring elevated desire.}
    \item \postreviewedit{\textbf{Memory-based override.} The injected memory document summarizes multiple low-composite entries and characterizes a recurring coping pattern. The model uses that summary to predict low ER~desire.}
\end{enumerate}

\postreviewedit{This selected entry matches all four binary labels and shows how historical outcomes appear in model context; the rationale itself is not a mechanism test.}

\subsection{Selected Rationale: Two Historical-Outcome Channels.}
\label{sec:paa-case-frazzled}

\postreviewedit{This selected case illustrates two historical information channels available to the model.}

\postreviewedit{Participant~258's profile records high trait negative affect and moderate depression. On an evening EMA, the diary reads: \emph{``Tired and a bit frazzled not sure why.''} The model's rationale cites two types of evidence:}

\postreviewedit{\textbf{Signal~1: Same-user labeled history.} Three labeled entries are cited in the rationale. The model groups diary text as internally or externally focused and associates those groups with the EMA-derived composite. It also receives 35.7~km of travel ($z=+1.79$ relative to the tool's baseline). These are model-generated interpretations from a very small sample, not verified coping or receptivity patterns.}

\postreviewedit{\textbf{Signal~2: Peer outcomes.} \path{find_peer_cases} returns 15 diary cases containing similar ``tired'' or ``frazzled'' language and their affect and ER labels. The model cites the retrieved label distribution and participant profile when producing its prediction. This example does not test the effect of retrieval.}

\clearpage
\reviewrevision{\section{Participant Demographics}\label{app:demographics}}

\begin{table}[H]
\centering
\color{blue}
\caption{\reviewrevision{Demographics of the 50 evaluation participants (BUCS baseline survey).}}
\label{tab:app-demographics}
\small
\begin{tabular}{ll}
\toprule
\textbf{Characteristic} & \textbf{Value} \\
\midrule
Age, mean (SD), range & 53.5 (10.9), 30--77 \\
Gender & 45 female (90\%), 4 male (8\%), 1 other (2\%) \\
Race & 47 White (94\%); 1 Black; 1 American Indian/Alaska Native; 1 other \\
Ethnicity & 1 Hispanic (2\%); 1 missing \\
Cancer type & 28 breast (56\%); 6 other; 3 melanoma; 3 endometrial; 2 prostate; \\
 & 2 lung/bronchus; 1 each colorectal, bladder, kidney, pancreatic, \\
 & leukemia, non-Hodgkin lymphoma \\
Cancer stage & 10 stage I; 4 stage II; 5 stage III; 11 stage IV; 8 in-situ/0; 12 N/A \\
Education (years), median (range) & 16 (12--30) \\
Marital status & 33 married (66\%); 8 divorced; 7 single; 2 widowed \\
\bottomrule
\end{tabular}
\end{table}

\end{document}